\documentclass[aps,pre,preprint,showpacs,nofootinbib]{revtex4}

\begin{document}

\preprint{LPT-Orsay 06-38}

\title{Electric-field correlation in quantum charged fluids coupled to the
radiation field}

\author{B.Jancovici}
\email{Bernard.Jancovici@th.u-psud.fr}
\affiliation{Laboratoire de Physique Th\'eorique, Universit\'e de Paris-Sud, 
91405 Orsay Cedex, France\footnote{Unit\'e Mixte de Recherche No 8627 - CNRS}}

\date{\today}

\begin{abstract}
In a recent paper [S.El Boustani, P.R.Buenzli, and Ph.A.Martin, Phys.Rev. E
{\bf 73}, 036113 (2006)], about quantum charges in equilibrium with
radiation, among other things the asymptotic form of the electric-field
correlation has been obtained by a microscopic calculation. It has been found
that this correlation has a long-range algebraic decay of the form $1/r^3$
(except in the classical limit). The macroscopic approach, in the Course of
Theoretical Physics of Landau and Lifshitz, gives no such decay. In
this Brief Report, we revisit and complete the macroscopic approach of Landau
and Lifshitz, and suggest that, perhaps, the use of a classical electromagnetic
field by El Boustani et al. was not justified.  
\end{abstract}


\maketitle

\section{INTRODUCTION}

We consider the equilibrium quantum statistical mechanics of an infinite and
homogeneous system of nonrelativistic charged particles coupled to
electromagnetic radiation. For this system, we investigate the correlation
function of the electric field at equal times 
\mbox{$<E_i({\bf r}_1)E_j({\bf r}_2)>$} where $E_i({\bf r}_1)$ is
the quantum operator for the $i{\rm th}$ Cartesian component ($i=1,2,3$) of
the electric field at point ${\bf r}_1$ and $<\ldots>$ denotes a quantum 
statistical average at temperature $T$; in the present homogeneous medium,
this correlation function depends on ${\bf r}_1$ and ${\bf r}_2$ only through
${\bf r}={\bf r}_1-{\bf r}_2=(x_1,x_2,x_3)$. We want to compute the form of
this correlation function, when $r$ becomes large compared to the microscopic
scale. 

A macroscopic approach to this problem has been made a long time ago by
Landau and Lifshitz (LL) in their famous Course of Theoretical 
Physics\cite{LL,LP}. Actually, their theory has been written for any medium,
characterized by a complex frequency-dependent dielectric function
$\epsilon(\omega)$. Recently, a microscopic theory has been elaborated by El
Boustani, Buenzli, and Martin (BBM)\cite{BBM}. They find an electric-field
correlation function in disagreement with the one advocated by LL: while BBM
find that the correlation function has a long-range power-law decay of the
form $1/r^3$ (except in the classical limit), I could not extract this
algebraic decay from the work of Landau and Lifshitz. The reason for this
disagreement is an open problem. 

As a first step for clarifying this problem, the present paper revisits and
completes the macroscopic approach of LL. Section II
summarizes  this approach, making it more explicit about the decay of the
electric-field correlation function. In Section III, the formalism is applied
to the special case of a one-component plasma. Section IV is a
Conclusion where the above-mentioned discrepancy is discussed.

\section{MACROSCOPIC APPROACH}

\subsection{The results of Landau and Lifshitz}

LL have solved the macroscopic Maxwell equations, in a medium
characterized by a complex frequency-dependent dielectric function
$\epsilon(\omega)$, in presence of a random field. This macroscopic approach
is expected to be valid only for distances large compared to the microscopic
scale. The fluctuation-dissipation theorem is used for obtaining the
electric-field correlation as a Fourier transform with respect to ${\bf r}$
and the time difference (here this time difference is zero): 
\begin{equation}
<E_i({\bf r}_1)E_j({\bf r}_2)>=\int_{-\infty}^{\infty}
\frac{{\rm d}\omega}{2\pi}\int\frac{{\rm d}^3k}{8\pi^3}
\exp({\rm i}{\bf k}\cdot{\bf r})\mathcal{E}_{ij}({\bf k},\omega)
\label{FT1}
\end{equation}
where, in terms of the wave vector ${\bf k}$ and the frequency $\omega$ (see 
\S 76 and 77 of \cite{LP}),
\begin{equation}
\mathcal{E}_{ij}({\bf k},\omega)=
-4\pi\hbar\coth\frac{\hbar\omega}{2T}
{\rm Im}\frac{\omega^2/c^2}{(\omega^2/c^2)\epsilon(\omega)-k^2}
\left[\delta_{ij}-\frac{c^2}{\omega^2\epsilon(\omega)}k_ik_j\right];
\label{FT2}  
\end{equation}
$\hbar$ is Plank's constant divided by $2\pi$, $c$ is the velocity of light
in vacuum, $T$ is the temperature in units of energy.

In order to make easier the comparison with the paper by BBM, we shall now
separate (\ref{FT2}) in its longitudinal and transverse parts. 

\subsection{Longitudinal and transverse correlations}

Introducing the projectors $k_ik_j/k^2$ and $\delta_{ij}-(k_ik_j/k^2)$ on
the longitudinal and transverse parts of the electric-field correlation 
functions, we can rewrite (\ref{FT2}) as 
\begin{equation}
\mathcal{E}_{ij}({\bf k},\omega)=\mathcal{E}_{ij}^{{\rm l}}({\bf k},\omega)
+\mathcal{E}_{ij}^{{\rm t}}({\bf k},\omega), \label{LT}
\end{equation}
where the longitudinal part is 
\begin{equation}
\mathcal{E}_{ij}^{{\rm l}}({\bf k},\omega)=
-4\pi\hbar\coth\frac{\hbar\omega}{2T}
{\rm Im}\frac{1}{\epsilon(\omega)}\frac{k_ik_j}{k^2} \label{l}
\end{equation}
and the transverse part is
\begin{equation}
\mathcal{E}_{ij}^{{\rm t}}({\bf k},\omega)=
-4\pi\hbar\coth\frac{\hbar\omega}{2T}
{\rm Im}\frac{\omega^2/c^2}{(\omega^2/c^2)\epsilon(\omega)-k^2}
\left(\delta_{ij}-\frac{k_ik_j}{k^2}\right). \label{t}
\end{equation}
It should be noted that the cross correlation between the longitudinal
and transverse parts of the electric field vanishes. Indeed, in Fourier
space, in terms of the Fourier transforms of these fields  
${\bf E}^{\rm l}({\bf k},\omega)$ and ${\bf E}^{\rm t}({\bf k},\omega)$, the
cross correlation tensor is proportional to
$<{\bf E}^{\rm l}({\bf k},\omega){\bf E}^{\rm t}(-{\bf k},-\omega)>$, where
${\bf E}^{\rm l}$ is along ${\bf k}$ and ${\bf E}^{\rm t}$ is normal to
${\bf k}$. Since the medium is isotropic, the correlation tensor is unchanged
if ${\bf E}^{\rm t}$ is replaced by its opposite; therefore, the correlation
tensor is equal to its opposite, i.e. it vanishes.

The integral on $\omega$ of (\ref{l}) converges: in particular, near
$\omega=0$, the dielectric function $\epsilon(\omega)$ behaves like 
$4\pi {\rm i}\sigma/\omega$, where $\sigma$ is the static
conductivity\cite{LL} (see Section III for the special case of a
one-component plasma) and (\ref{l}) is finite at $\omega=0$. The Fourier
transform of $4\pi/k^2$ is $1/r$ and the longitudinal part of (\ref{FT1}) is
\begin{equation} 
<E_i({\bf r}_1)E_j({\bf r}_2)>_{{\rm l}}\:= 
\left(\int_{-\infty}^{\infty}\frac{{\rm d}\omega}{2\pi}
\hbar\coth\frac{\hbar\omega}{2T}{\rm Im}\frac{1}{\epsilon(\omega)}\right)
\frac{\partial^2}{\partial x_i\partial x_j}\frac{1}{r}. \label{la}
\end{equation}
  
At small $k$, in (\ref{t}), the term 
$(\omega^2/c^2)/[(\omega^2/c^2)\epsilon(\omega)-k^2)]$ can be replaced by 
$1/\epsilon(\omega)$. Indeed, if $\omega\neq 0$, this term can be expanded in
powers of $k^2$ and $1/\epsilon(\omega)$ is the leading term. If
$|\omega|<\omega_0$, where $\omega_0$ is a sufficiently small constant,  
$\epsilon(\omega)=4\pi {\rm i}\sigma/\omega$,
$\hbar\coth\frac{\hbar\omega}{2T}=2T/\omega$, and
\begin{equation}
\int_{-\omega_0}^{\omega_0}\frac{{\rm d}\omega}{2\pi}
\mathcal{E}_{ij}^{{\rm t}}({\bf k},\omega)=
\frac{2T}{\pi\sigma}\left[\omega_0-\frac{(ck)^2}{4\pi\sigma}
\arctan\frac{4\pi\sigma \omega_0}{(ck)^2}\right]
\left(\delta_{ij}-\frac{k_ik_j}{k^2}\right). \label{t1}
\end{equation}
At small $k$, the $\arctan$ in (\ref{t1}) behaves like $\pi/2$. Finally, for
all values of $\omega$, at small $k$,
\begin{equation}
\mathcal{E}_{ij}^{{\rm t}}({\bf k},\omega)\sim
-4\pi\hbar\coth\frac{\hbar\omega}{2T}\left[\left({\rm Im}\frac{1}
{\epsilon(\omega)}\right)+O(k^2)\right]
\times\left(\delta_{ij}-\frac{k_ik_j}{k^2}\right). \label{t2}
\end{equation}
 
The asymptotic behavior of the transverse part of (\ref{FT1}) is given by the
most singular part at small $k$ of (\ref{t2}), which is just
opposite to (\ref{l}), i.e. the asymptotic form of  
$<E_i({\bf r}_1)E_j({\bf r}_2)>_{{\rm t}}$ is opposite to (\ref{la}).
These asymptotic behaviors of the form $1/r^3$ cancel each other in the total
correlation function (\ref{FT1}). This cancellation had been previously
noted\cite{F,F1} in the  classical limit $T\rightarrow\infty$, but the present
macroscopic approach predicts that this cancellation persists in the quantum
regime at any temperature, contrarily to the prediction of the microscopic
theory of BBM. 

The integral on $\omega$ in (\ref{la}) is simply related to the second moment
of the charge correlation function. Indeed the charge density $\rho$ is given
by the Poisson equation ${\rm div}{\bf E}=4\pi\rho$. Only the longitudinal
part of ${\bf E}$ has a non-vanishing divergence. Therefore, since the Fourier
transform (\ref{l}) of $<E_i({\bf r}_1)E_j({\bf r}_2)>_{{\rm l}}$ is 
of the form  $Ak_ik_j/k^2$, the Fourier tranform of  
$<\rho({\bf r}_1)\rho({\bf r}_2)>$ is $(4\pi)^{-2}Ak^2$, which means
that $A$ is $-(3\pi)^{-1}$ times the second moment of this charge correlation
function. Taking the inverse Fourier transform of $Ak_ik_j/k^2$ gives for the
longitudinal field correlation 
\begin{equation}
<E_i({\bf r}_1)E_j({\bf r}_2)>_{{\rm l}}\:= 
-\frac{\partial^2}{\partial x_i\partial x_j}\frac{1}{r}
\left[-\frac{2\pi}{3}
\int{\rm d}^3r'\:r'^2<\rho({\bf 0})\rho({\bf r'})>\right], \label{2m} 
\end{equation}
for $r$ large compared to the microscopic scale, in agreement with a
microscopic derivation\cite{LM,PM}. 

In the classical limit $T\rightarrow\infty$, 
$\hbar\coth\frac{\hbar\omega}{2T}\sim 2T/\omega$ and the integral on $\omega$
in (\ref{la}) (which also occurs in the transverse part with the opposite
sign) has the simple value $-T$ \cite{F}. This can be shown\cite{F2} by
invoking that $\epsilon(\omega)$ has no zeros when $\omega$ is in the complex
upper half-plane. The calculation is made in the Appendix. Therefore, in the
classical limit, 
\begin{equation}
<E_i({\bf r}_1)E_j({\bf r}_2)>_{{\rm l}}\sim\:
-T\frac{\partial^2}{\partial x_i\partial x_j}\frac{1}{r}, \label{lac}     
\end{equation}
and
\begin{equation}
<E_i({\bf r}_1)E_j({\bf r}_2)>_{{\rm t}}\sim\:
T\frac{\partial^2}{\partial x_i\partial x_j}\frac{1}{r}; \label{tac}
\end{equation}
these classical results can also be obtained from (\ref{2m}) since, in the
classical case, the second moment in (\ref{2m}) obeys the Stillinger-Lovett
sum rule\cite{sl}
\begin{equation}
-\frac{2\pi}{3}\int{\rm d}^3r'\:r'^2<\rho({\bf 0})\rho({\bf
 r'})>=T. \label{SL}
\end{equation}
The asymptotic classical transverse correlation function (\ref{tac}) is the
one of a free field. This is in agreement with the Bohr-van Leeuwen theorem
\cite{B,vL,AA} which says that, in equilibrium classical statistical
mechanics, matter and radiation are uncoupled.

\section{ONE-COMPONENT PLASMA}

\subsection{Drude dissipationless dielectric function}

The above formalism simplifies in the special case of a one-component plasma,
a system of one species of particles of charge $e$, mass $m$, and number
density $n$, in a neutralizing homogeneous background. We are interested in
small wave numbers, when the dissipation goes to zero\cite{NP} (i.e. the
static conductivity is infinite), and the dielectric function can be taken as
the Drude one  
\begin{equation}
\epsilon(\omega)=1-\frac{\omega_{{\rm p}}^2}{\omega(\omega+{\rm i}\eta)},
\label{D}
\end{equation}
where $\omega_{\rm p}=(4\pi ne^2/m)^{1/2}$ is the plasma frequency and the
dissipation constant $\eta$ is taken as infinitesimal. 

Then, using ${\rm Im}[1/\epsilon(\omega)]=
-\pi (\omega/|\omega|)\omega_{{\rm p}}^2
\delta(\omega^2-\omega_{{\rm p}}^2)$ in (\ref{l}), one finds for the
longitudinal correlation function  
\begin{equation}
<E_i({\bf r}_1)E_j({\bf r}_2)>_{{\rm l}}\:=
-\frac{1}{2}\hbar\omega_{{\rm p}}\coth\frac{\hbar\omega_{{\rm p}}}{2T}\:
\frac{\partial^2}{\partial x_i\partial x_j}\frac{1}{r}. \label{lD}
\end{equation}

In a  similar calculation, using (\ref{D}) in (\ref{t}), 
$\delta(\omega^2-\omega_{{\rm p}}^2-c^2k^2)$ appears, and one finds for the
transverse correlation function
\begin{eqnarray}
<E_i({\bf r}_1)E_j({\bf r}_2)>_{{\rm t}}&=&
\int\frac{{\rm d}^3k}{8\pi^3}\exp({\rm i}{\bf k}\cdot{\bf r})
\frac{1}{2}\hbar(\omega_{{\rm p}}^2+c^2k^2)^{1/2} \nonumber  \\
&\times&\coth\frac{\hbar(\omega_{{\rm p}}^2+c^2k^2)^{1/2}}{2T}\:4\pi
\left(\delta_{ij}-\frac{k_ik_j}{k^2}\right). \label{tD}
\end{eqnarray}
For small $k$, the $k$-singularity in (\ref{tD}) comes from the term
$k_ik_j/k^2$ with the replacement of 
$(\omega_{{\rm p}}^2+c^2k^2)^{1/2}$ by $\omega_{{\rm p}}$. Then, one finds
for the asymptotic form of the transverse correlation function just the
opposite of the longitudinal one (\ref{lD}).

\subsection{Longitudinal and transverse modes}

The above results for the electric-field correlation functions can also be
obtained, perhaps in a more transparent way, from the modes of vibration of
the plasma. In this subsection, we shall not use the dielectric function 
$\epsilon(\omega)$, but rather take into account explicitly in the equations
the charge and electric-current densities. 

For every wave vector ${\bf k}$ there is a longitudinal mode; its
frequency\cite{J} is $\omega_{{\rm p}}$ (neglecting a term of order $k^2$,
which is consistent with the previous use of an $\epsilon$ independent of
${\bf k}$). In such a mode, the electric field, at position ${\bf r}$ and 
time $t$ is of the form 
\begin{equation}
{\bf E}_{{\bf k}}({\bf r},t)= {\rm Re}({\bf k}/k)E_0
\exp({\rm i}{\bf k}\cdot{\bf r}-{\rm i}\omega_{{\rm p}}t). \label{Ek}
\end{equation}
Since this mode is an oscillator, for studying it in quantum mechanics, we
can first use classical mechanics, and quantize at the end.
A collective velocity ${\bf v}_{{\bf k}}$ is given by Newton's law:
\begin{equation}
m\frac{{\rm d}{\bf v}_{{\bf k}}}{{\rm d}t}=
e{\bf E}_{{\bf k}}. \label{vk}
\end{equation}
From (\ref{vk}) one easily finds that the temporal average of the kinetic
energy density associated to ${\bf v}_{{\bf k}}$ is equal to the temporal
average of the energy density of the electric field. Taking into account that
the temporal average of the squared real electric field is $|E_0|^2/2$ and
equating the temporal average of the total energy to the statistical average
value for an oscillator (including the zero-point energy) gives   
\begin{equation}
\frac{|E_0|^2}{8\pi}=\frac{1}{V}
\frac{\hbar\omega_{{\rm p}}}{2}\coth\frac{\hbar\omega_{{\rm p}}}{2T}, 
\label{energyl}
\end{equation}
where $V$ is the volume of the system. The  contribution of this mode
(\ref{Ek}) to the electric-field correlation is 
${\rm Re}(|E_0|^2/2)\exp({\rm i}{\bf k}\cdot{\bf r})(k_ik_j/k^2)$, where
$|E_0|^2$ is given by (\ref{energyl}). Summing on all ${\bf k}$ wave vectors,
i.e. computing the integral $V\int{\rm d}^3k/(8\pi^3)\ldots$ reproduces
(\ref{lD}). 

For every wave vector ${\bf k}$, there are also two (there are two possible
polarizations) transverse modes of frequency\cite{J}
$\omega_k=(\omega_{{\rm p}}^2+c^2k^2)^{1/2}$. For such a mode, the
electric field is of the form
\begin{equation}
{\bf E}_{{\bf k}}({\bf r},t)= {\rm Re}\:{\bf e}E_0
\exp({\rm i}{\bf k}\cdot{\bf r}-{\rm i}\omega_kt), \label{Ekt}
\end{equation}
where ${\bf e}$ is a unit vector normal to ${\bf k}$, and the
magnetic-induction field is 
\begin{equation}
{\bf B}_{{\bf k}}({\bf r},t)=
\frac{c}{\omega_k}{\bf k}\times{\bf E}_{{\bf k}}. \label{Bkt}
\end{equation}
Again, a collective velocity is given by Newton's law (\ref{vk}) (the 
magnetic force can be neglected for non-relativistic velocities).The 
temporal average of the electric-field energy density is $|E_0|^2/(16\pi)$. 
The temporal average of the  kinetic energy density is 
$(\omega_{{\rm  p}}^2/\omega_k^2)|E_0|^2/(16\pi)$, from (\ref{vk}). The
temporal average of the magnetic-induction energy density is  
$(c^2k^2/\omega_k^2)|E_0|^2/(16\pi)$, from (\ref{Bkt}). Thus, the temporal
average of the total energy density again is $|E_0|^2/(8\pi)$, and
\begin{equation}
\frac{|E_0|^2}{8\pi}=
\frac{1}{V}\frac{\hbar\omega_k}{2}\coth\frac{\hbar\omega_k}{2T}.  
\label{energyt}
\end{equation}
The contribution of the transverse mode (\ref{Ekt}) to the electric-field
correlation is 
${\rm Re}\:e_ie_j(|E_0|^2/2)\exp({\rm i}{\bf k}\cdot{\bf r})$ where
$|E_0|^2$ is given by (\ref{energyt}). Adding the contribution of the other
polarization replaces $e_ie_j$ by $\delta_{ij}-(k_ik_j/k^2)$. Summing on all
${\bf k}$ wave vectors reproduces (\ref{tD}).

\section{CONCLUSION}

The macroscopic approach of LL is in agreement with the
microscopic calculation of BBM, for the longitudinal part of
the electric-field correlation (incidentally, this agreement is an indication
that the macroscopic approach can be correct). The disagreement is about the
transverse part only, which exactly cancels (\ref{la}) in the macroscopic
approach while it obeys (\ref{tac}) in the microscopic approach\cite{BBM},
even in the quantum regime. I am tempted to believe that the cancellation
predicted by the macroscopic calculation is essentially correct. Here are my
reasons. 

BBM\cite{BBM} point out that the macroscopic theory of LL uses a local
dielectric function $\epsilon(\omega)$ rather than a ${\bf k}$-dependent
one. Indeed it is tempting to use a longitudinal 
$\epsilon_{{\rm l}}(k,\omega)$ in (\ref{l}) and a transverse 
$\epsilon_{{\rm t}}(k,\omega)$ in (\ref{t}). However, the leading (singular) 
term of (\ref{l}) or (\ref{t}) at small $k$ would be still obtained by taking
these $k$-dependent dielectric functions at $k=0$ where both of them become
$\epsilon(\omega)$. Thus, a singular term\cite{Cornu} in the small-$k$
expansion of $\epsilon_{{\rm l}}(k,\omega)$ and/or 
$\epsilon_{{\rm t}}(k,\omega)$ would not change the prefactors of the terms
$k_ik_j/k^2$ (this cancellation only concerns the asymptotic terms of the 
form $1/r^3$, however an algebraic decay faster than $1/r^3$ could remain).
BBM make another criticism of the macroscopic approach of LL: it
neglects the magnetic permeability. It would be strange that taking the
magnetic permeability into account in the macroscopic approach would change
the transverse correlation function into the one of a free field (however, it
might bring small corrections to the results of LL\cite{F3}).

The final remark by BBM might be the key point: in their
microscopic approach, they use a \emph{classical} electromagnetic field. That
their asymptotic transverse correlation is the one of a free field,
decoupled from matter, even in the case of quantum particles, might be due to
this feature of their calculation. In III.B, I have argued that the
transverse modes certainly feel the presence of matter. Perhaps the argument,
in the 3rd paragraph of the Introduction of BBM, in favor of using a
classical electromagnetic field, has a flaw: The frequency of a transverse
mode is always larger than the plasma frequency $\omega_{{\rm p}}$, and the
condition $\beta\hbar ck\ll 1$ does \emph{not} imply that
$\beta\hbar\omega_{{\rm p}}\ll 1$. 

The microscopic calculation of BBM is very elaborate, cleverly using an
elegant path-integral formalism. It would certainly be very interesting to
redo this microscopic calculation with a quantized electromagnetic field, if
feasible.
 
\appendix*
\section{}
This Appendix is based on a private message from B.U.Felderhof. In the
classical limit, the integral in (\ref{la}) becomes $TI$ where $I$ is
\begin{equation}
I=\frac{1}{\pi}\int_{-\infty}^{\infty}\frac{{\rm d}\omega}{\omega}
{\rm Im}\frac{1}{\epsilon(\omega)}.  \label{int}
\end{equation}
We want to show that $I=-1$.

Since $\epsilon^{\ast}(\omega)=\epsilon(-\omega^{\ast})$, $I$ can be 
rewritten as
\begin{equation}
I=\frac{1}{{\rm i}\pi}\int_{-\infty}^{\infty}\frac{{\rm d}\omega}{\omega}
\frac{1}{\epsilon(\omega)}.   \label{int2}
\end{equation}
Since $\epsilon(\omega)$ has no zeros in the upper complex
half-plane\cite{LL}, the integral along the real axis in (\ref{int2}) can be
changed into an integral along the half-circle $C$ at infinity in the upper
half-plane. Since $\epsilon$ is 1 at infinity, this integral is 
$\int_C{\rm d}\omega/\omega =-{\rm i}\pi$. Therefore, $I=-1$.

The same reasoning applies to the transverse part. Comparing (\ref{l}) and
(\ref{t}), one sees that, on the half-circle at infinity, $1/\epsilon$ and 
$(\omega^2/c^2)/[(\omega^2/c^2)\epsilon-k^2]$ have the same limit 1. 
Furthermore, the term $k_ik_j/k^2$ in (\ref{l}) and the term 
$\delta_{ij}-(k_ik_j/k^2)$ in (\ref{t}) will have opposite inverse Fourier
transforms for $r\neq 0$. Thus, (\ref{tac}) is just opposite to (\ref{lac}).

\begin{acknowledgments}
I have benefited from fruitful discussions with P.R.Buenzli, Ph.A.Martin, and
B.U.Felderhof. 
\end{acknowledgments}

\end{document}